# Forensic Investigation of Social Media and Instant Messaging Services in Firefox OS: Facebook, Twitter, Google+, Telegram, OpenWapp and Line as Case Studies

**Mohd Najwadi Yusoff**
School of Computer Science, Universiti Sains Malaysia, Penang, Malaysia.
mohd.najwadi@gmail.com

**Ali Dehghantanha**
School of Computing, Science and Engineering, University of Salford, Manchester, United Kingdom.
A.Dehghantanha@Salford.ac.uk

**Ramlan Mahmod**
Faculty of Computer Science and Information Technology, Universiti Putra Malaysia, Serdang, Malaysia.
ramlan@upm.edu.my

## Abstract

Mobile devices are increasingly utilized to access social media and instant messaging services, which allow users to communicate with others easily and quickly. However, the misuse of social media and instant messaging services facilitated conducting different cybercrimes such as cyber stalking, cyber bullying, slander spreading and sexual harassment. Therefore, mobile devices are an important evidentiary piece in digital investigation. In this chapter, we report the results of our investigation and analysis of social media and instant messaging services in Firefox OS. We examined three social media services (Facebook, Twitter and Google+) as well as three instant messaging services (Telegram, OpenWapp and Line). Our analysis may pave the way for future forensic investigators to trace and examine residual remnants of forensics value in FireFox OS.

# 1. Introduction

The exponential growth of social media and instant messaging applications facilitated development of many serious cybercrime and malicious activities (Mohtasebi and Dehghantanha, 2011a). Cybercriminals are constantly changing their strategies to target rapidly growing social media and instant messaging users. The misuse of social media and instant messaging in mobile devices may allow cybercriminals to utilize these services for malicious purposes (Mohtasebi and Dehghantanha, 2011b) such as spreading malicious codes, obtaining and disseminating confidential information etc. Many social media and instant messaging providers have extended their services to mobile platforms (Dezfouli et al., 2015) which worsen the situation as users are in danger of losing even more private information (Taylor et al., 2012). Copyright infringement, cyber stalking, cyber bullying, slander spreading and sexual harassment are becoming serious threats to social media and instant messaging mobile users (Dezfouli et al., 2013). Therefore, it is common to confront with different types of mobile devices during variety of forensics investigation cases (Damshenas et al., 2014). Mobile devices are now an important source of forensic remnants relevant to users social media and instant messaging activities (Mohtasebi and Dehghantanha, 2013). However, difference between mobile devices mandate forensics investigators to develop customized methods and techniques for investigation of different phones (Mohtasebi et al., 2012).

With the emergence of smartphones, almost all parts of phones such as internal storage, flash memory and internal volatile memory contains valuable evidences (Yang et al., 2016). Chen extracted SMS, phone book, call recording, scheduling, and documents from Windows Mobile OS via Bluetooth, Infrared and USB using Microsoft ActiveSync (Chen et al., 2009). The acquired data were extracted from mobile phone internal storage, SIM card as well as removable memory. Irwin and Hunt have successfully mapped internal and external memory of Windows Mobile ver.5 running on IPAQ Pocket PC over wireless connections using their own developed forensic tools (Irwin and Hunt, 2009). Pooters has developed a forensic tool called Symbian Memory Imaging Tool (SMIT) to create bit-by-bit copies of the internal flash memory of Symbian OS phones such as Nokia E65, E70 and N78 model (Pooters, 2010). Lessard and Kessler (Lessard and Kessler, 2010) have acquired a bit-by-bit logical image of a HTC Hero memory using UNIX dd command and analyzed the resulted images using AccessData Forensic Toolkit (FTK) Imager v2.5.1 (Accessdata, 2007). Gómez-Miralles and Arnedo-Moreno have utilized a Jailbroken iPad's camera connection kit to acquire an image of the device via USB connection (Gómez-Miralles and Arnedo-Moreno, 2012). Iqbal has enhanced Gómez-Miralles and Arnedo-Moreno method by developing a method to acquire iOS memory images without jail-breaking the device (Iqbal et al., 2012). Sylve has presented a methodology and toolset for acquisition volatile physical memory of Android devices by creating a new kernel module for dumping the memory (Sylve et al., 2012).

Beyond evidence acquisition, many researchers have shown big interest in investigating social media and instant messaging services on different mobile platforms. Husain and Sridhar have analyzed AIM, Yahoo! Messenger and Google Talk instant messaging applications in Apple iOS to detect potential application of these instant messaging services, particularly in cyber bullying and cyber stalking (Husain and Sridhar, 2010). They have managed to detect username, password, buddy list, last login time and conversation contents together with timestamp. Jung has analyzed eight social media applications in Apple iOS namely Cyworld, Me2Day, Daum Yozm, Twitter, Facebook, NateOn UC., KakaoTalk and MyPeople (Jung et al., 2011) and managed to retrive user info, friend list, message, contact and media informtion of each application. Tso has observed the diversification of the backup files for Facebook, WhatsApp Messenger, Skype, Windows Live Messenger and Viber in Apple iOS (Tso et al., 2012). Anglano has analyzed WhatsApp Messenger application remnants on an Android smartphone and reconstructed the list of contacts and the chronology of communicated messages (Anglano, 2014). Karpisek has successfully decrypted the network traffic of WhatsApp Messenger and managed t obtain forensic artifact related to call features and visualized messages that have been exchanged between users (Karpisek et al., 2015). Walnycky has examined 20 popular instant messaging applications on Android platform and has reconstructed some or the entire message content of 16 applications (Walnycky et al., 2015). Said has conducted a comparative study of Facebook and Twitter remnants on Apple iOS, Windows Mobile and RIM BlackBerry (Said et al., 2011) and extracted Facebook and Twitter remnants of Apple iOS, and Facebook. Mutawa has compared evidences of Facebook, Twitter and MySpace in three different operating systems namely Apple iOS, Google Android and RIM BlackBerry and could not recover any artifact from Blackberry while iPhones and Android contained many valuable artefacts (Al Mutawa et al., 2012). Iqbal (Iqbal et al., 2014) has compared artifacts of Samsung's ChatON application between a Samsung Galaxy Note running Android 4.1 and an Apple iPhone running with iOS 6 and managed to detect all sent and received messages with timestamp and location of the sent files on both platforms. Dezfouli investigated Facebook, Twitter, LinkedIn and Google+ on Android and Apple iOS platforms and managed to recover many artifacts including username, contact information, location, friend list, social media post, messages, comments and IP addresses of selected social media applications on both platforms (Dezfouli et al., 2015). Table 1 is summarizing the literature on social media and instant messaging investigation forensics.

Table 1 - Summary of Social Media and Instant Messaging Investigation on Multiple Mobile Platforms

| Researcher(s) | Application(s) | Platform(s) |
|---|---|---|
| **Husain and Sridhar 2010** | AIM, Yahoo! Messenger, Google Talk | Apple iOS |
| **Jung et al. 2011** | Cyworld, Me2Day, Daum Yozm, Twitter, Facebook, NateOn UC., KakaoTalk, MyPeople | Apple iOS |
| **Tso et al. 2012** | Facebook, WhatsApp Messenger, Skype, Windows Live Messenger, Viber | Apple iOS |
| **Anglano 2014** | WhatsApp Messenger | Google Android |
| **Karpisek et al. 2015** | WhatsApp Messenger | Google Android |
| **Walnycky et al. 2016** | WhatsApp Messenger, Viber, Instagram, Okcupid, ooVoo, | Google Android |

| | Tango, Kik, Nimbuzz, eetMe, MessageMe, TextMe, Grindr, HeyWire, Hike, textPlus, Facebook Messenger, Tinder, Wickr, Snapchat, Blackberry Messenger | |
|---|---|---|
| **Said et al. 2011** | Facebook, Twitter | Apple iOS, Windows Mobile, RIM BlackBerry |
| **Al Mutawa et al. 2012** | Facebook, Twitter, MySpace | Apple iOS, Google Android, RIM Blackberry |
| **Iqbal et al. 2014** | Samsung's ChatON | Apple iOS, Google Android |
| **Dezfouli et al. 2015** | Facebook, Twitter, LinkedIn, Google+ | Apple iOS, Google Android |

As can be seen from Table 1, there was no previous work investigating remnants of Facebook, Twitter, Google+, Telegram, OpenWapp and Line applications on Firefox OS (FxOS) which is the gap targeted in this paper. In this chapter, we present investigation of a Geeksphone, model name Peak (Geeksphone, 2013) running FxOS 1.1.1 as the main subject of the investigation. Two binary images were taken from the phone internal storage and internal memory and then valuable forensics remnants were examined. We have mainly interested in evidential remnants of user activities with Facebook, Twitter, Google+, Telegram, OpenWapp and Line applications.

The rest of this chapter is organized as follows. In section 2, we will explain the methodology used in our research and in section 3, we will outline the setup for our experiment. In section 4, we will present our research findings and conclude our research in section 5.

## 2. Methodology

This research has performed an investigation on Mozilla FxOS running on a phone released by Geeksphone, called Peak (Geeksphone, 2013). Released in April 2013, this phone is equipped with FxOS version 1.1.1. Table 2 shows the full specification of Geeksphone Peak.

Table 2 - Geeksphone Peak Full Specification

| Hardware | Detail |
|---|---|
| **Processor** | 1.2 GHz Qualcomm Snapdragon S4 8225 processor (ARMv7) |
| **Memory** | 512 MB Ram |
| **Storage** | -Internal 4GB<br>-Micro SD up to 16GB |
| **Battery** | - 1800 mAh<br>- micro-USB charging |
| **Data Inputs** | Capacitive multi-touch IPS display |
| **Display** | 540 × 960 px (qHD) capacitive touchscreen, 4.3" |
| **Sensor** | -Ambient light sensor<br>-Proximity sensor<br>-Accelerometer |
| **Camera** | 8 MP (Rear), 2 MP (Front) |
| **Connectivity** | -WLAN IEEE 802.11 a/b/g/n |
| | -Bluetooth 2.1 +EDR |
| | -micro-USB 2.0 |
| | -GPS |
| | -mini-SIM card |
| | -FM receiver |
| **Compatible Network** | - GSM 850 / 900 / 1800 / 1900 |
| | - HSPA (Tri-band) |
| | - HSPA/UMTS 850 / 1900 / 2100 |
| **Dimension** | -Width: 133.6 millimetres (5.26 in) |
| | -Height: 66 millimetres (2.6 in) |
| | -Thickness: 8.9 millimetres (0.35 in) |

As illustrated in the Figure 1, initially the device setting were wiped and restored to the default factory setting. The acquisition process was then executed to acquire FxOS phone image (.ffp) and memory image (.ffm). The first two binary images were then marked as base image and their MD5 hash values were preserved. Next, the investigator installed social media or instant messaging applications to the phone via Mozilla Market Place. The investigator simulate the actual use of the application by running communication activities such as posting, sending private message, received reply, upload picture, mentioning, following, and many more social media and instant messaging activities. Each of detail steps, credentials and communication activities were documented properly in forensically sound manner. Finally, second acquisition process performed to identify and investigate what artifact are likely to remain, what type of credential can be extracted, and what data remnants could be recovered. Both images were named according to the type of application installed and its MD5 hash values were taken. This step meant for comparison to the earlier data acquisition, to see if there were any differences with the other MD5 hash values.

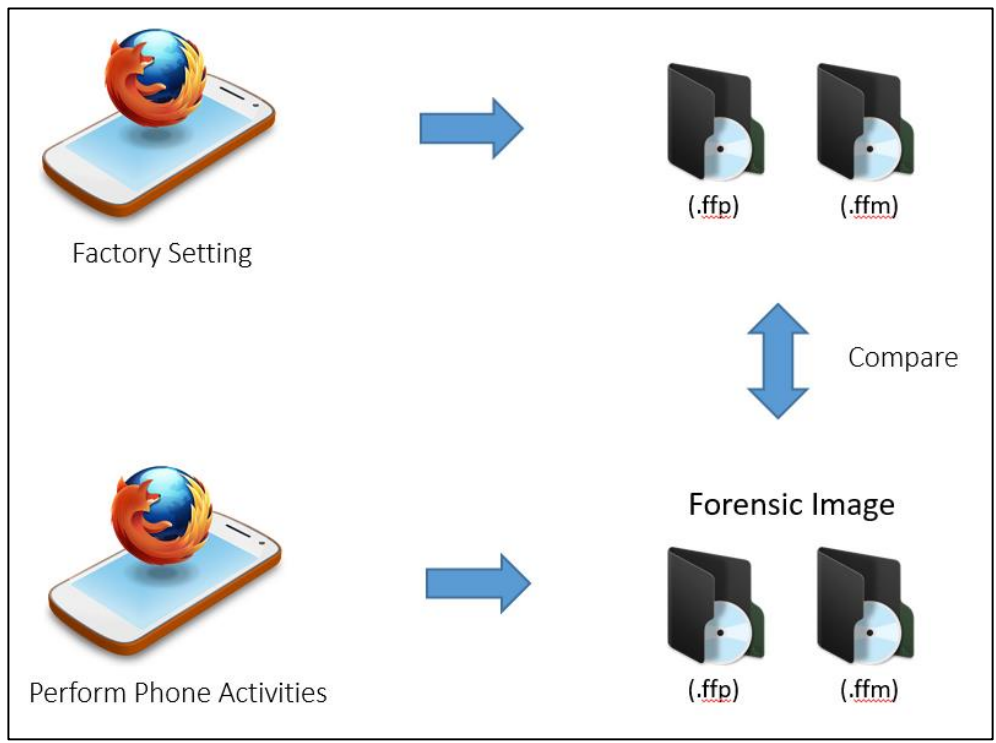

Figure 1 - Method to Create Forensic Evidences

Our method to create forensic evidences in Figure 1 were performed by using Facebook, Twitter, Telegram, OpenWapp and Line applications. We also repeat this method using Facebook, Twitter, Google+ and Telegram mobile web, in order to compare both FxOS applications artifacts and mobile web activities artifacts. However, comparison was not possible for OpenWapp and Line application artifacts, since both applications were not offered in mobile web platforms. It is vital to include the mobile web activities in this investigation since FxOS application were designed based on web application. Therefore, this research was attempting to proof that the investigation result of web-centric OS are totally different with other mobile OS investigation.

## 3. Experiment Setup

The research experiment was divided into five stages; (1) Preparing the host machine for acquisition process and analysis work; (2) Acquisition of phone image and memory image; (3) Installing the phone with targeted application; (4) Executed the activities and documenting all steps taken; and (5) Comparing base image with activities images

### 3.1 Preparing the Host Machine for Acquisition Process and Analysis Work

The evidences were acquired using Ubuntu 14.04 LTS and then analyzed in Windows 8.1 machine. The need to use separated operating system is due to the fact that memory acquisition will only work on Linux environment; whereas most of the analysis tools were design under Windows operating system. To successfully capture the phone image using Ubuntu, the host machine need to be running under Android Debug Bridge (ADB) (Yusoff et al., 2014a, 2014b, 2014c, 2014d). Following command was used to configure ADB package in Ubuntu:

```
# sudo apt-get install android-tools-adb
```

As for the volatile memory acquisition, we configure Linux Memory Extractor (LiME) (Sylve et al., 2012) using the following command:

```
# sudo apt-get install lime-forensics-dkms
```

AccessData Forensic Toolkit (FTK) version 3.1.2 (Accessdata, 2007) and HxD Hex Editor 1.7.7.0 (Maël Hörz, 2009) were installed in Windows machine to analyze captured forensic images and QtADB 0.81 Android Manager (QtADB, 2011) was used to browse the system files in the Geeksphone Peak.

### 3.2 Acquisition of Phone Image and Memory Image

Two types of binary images were extracted, meant to be used as forensic evidence in our case studies. The first binary image were extracted from FxOS phone internal memory using dd command. FxOS phone was first connected to the host machine and then ADB connection was started before we proceed to execute dd command. We have used the following command to start ADB connection between the phone and the host machine.

```
# adb shell
```

Once the connection is established, we have performed the following dd command to copy bit-by-bit of phone internal memory into SD card and we named it as FxOS phone memory (.ffp):

```
# dd if=/dev/block/mmcblk0 of=/mnt/emmc/base.ffp bs=2048
```

The whole binary image of FxOS internal memory was then copied from SD card into the host machine and we named it as base.ffp. We used block size 2MB during dd command execution as per suggested in previous studies (Yusoff et al., 2014b).

The next step was the extraction of the second binary image from FxOS phone volatile memory using LiME. LiME were design as such to allow the acquisition of volatile memory from Linux-based devices and originally tested using Android phones (Sylve et al., 2012). ADB was then used to load LiME, and the way of volatile memory acquisition at this step is similar to the dd command. We used the following command in ADB to load LiME from phone SD card:

```
# adb push lime.ko /mnt/emmc/lime.ko
```

After LiME were loaded, we run insmod to copy live bit-by-bit of volatile memory into SD card. For volatile memory, we named it as FxOS memory (.ffm)

```
# insmod /mnt/emmc/lime.ko "path=/mnt/emmc/base.ffm format=lime"
```

Both internal phone images (.ffp) and memory images (.ffm) were directed to SD card. After the acquisition process has completed, we killed the ADB connection, and then mounted the SD card to copy both images. We set the same name for both images and only the extension and file size are left as the differences. The steps were repeated for every action in our experiment.

### 3.3 Installing the Phone with Targeted Application

The objective of this experiment is to investigate both social media network and instant messaging platform in FxOS. Due to the limited number of applications offered in Mozilla Market Place, we have only managed to investigate 5 applications and 4 mobile webs as shows in Table 3

Table 3 - Selected Applications and Mobile Web

| Group | Application | Mobile Web |
| --- | --- | --- |
| **Social Media** | 1. Facebook<br>2. Twitter | 1. Facebook<br>2. Twitter<br>3. Google+ |

| Instant Messaging | 1. Telegram<br>2. OpenWapp<br>3. Line | 1. Telegram |
|---|---|---|

The investigation were performed on both application and the mobile web. The purpose of investigating both applications and mobile web, were to list out the differences and similarities between FxOS applications with its mobile web.

## 3.4 Executed the Activities and Documenting All Steps Taken

Our investigation in this experiment has started the moment we installed the applications in to the phone. We have also created a few dummy accounts for the experiments. All detail steps, credentials and communication activities were properly documented. In order to facilitate the use of credential during our experiment, we created all social media account using the same email address and password which are "mohd.najwadi@gmail.com" and "najwadi87" respectively. In general, all social media and instant messaging application will have one extra step; which is installation of the application. In contrary for mobile web, we can just browse directly from the site. Table 4 shows all steps and activities taken for social media experiment

Table 4 - Detail Steps for Social Media Experiment

| Social Media | Steps and Action | Image Filename |
|---|---|---|
| **Facebook Application** | Installing Facebook Application | Facebook-Install |
| | Login Facebook Account (ID : Root Wadi)<br>- Email : mohd.najwadi@gmail.com<br>- Password : najwadi87 | Facebook-Login |
| | Facebook Activities<br>- Post Status : Posting with love<br>- PM Mohd Najwadi : Hi. Test send msg<br>- Received Reply : received with thanks | Facebook-Post |
| **Facebook Web** | Login Facebook thru web www.facebook.com<br>- Email - mohd.najwadi@gmail.com<br>- Password - najwadi87 | Facebook-Browse |
| | Facebook Activities<br>- Post : Posting from web<br>- PM Mohd Najwadi : Test again<br>- Received Reply : received second time | Facebook-xtvt |
| **Twitter Application** | Installing Twitter Application | Twitter-Install |
| | Login Twitter Account (ID : @wadieq)<br>- Email : mohd.najwadi@gmail.com<br>- Password : najwadi87 | Twitter-Login |
| | Twitter Activities<br>- Tweet : My first tweet<br>- Mention : @wadieq test mention<br>- Favourites : @wadieq Taman tepi rumah | Twitter-Tweet |
| **Twitter Web** | Login Twitter thru web www.twitter.com<br>- Email : mohd.najwadi@gmail.com<br>- Password : najwadi87 | Twitter-Browse |
| | Twitter Activities<br>- Tweet : My second tweet<br>- Reply : @wadieq Reply comment<br>- Unfollow @MunirRashid | Twitter-xtvt |
| **Google+ Web** | Login Google+ thru web plus.google.com | Google-Login |
|  | - Email : mohd.najwadi@gmail.com<br>- Password : najwadi87 |  |
|  | Google+ Activties<br>- Follow people : tomska, british motogp<br>- create circle : firefox plus<br>- post : google plus firefox<br>- comment : komen sendiri | Google-xtvt |

There were 12 time acquisition processes were performed for each internal phone image and memory image altogether; and the files were name according to the steps taken as shown in Table 4. We have also tested with Facebook and Twitter, both on applications and mobile web, while for Google+ we can only test it on web since there is no application supported for Google+ in Mozilla Market Place at the moment we conducted the experiment. The phones were restored back to their factory setting only after we want to start with new application or mobile web experiments. For social media investigation, we restored the phones 5 times, the number of experiment conducted. After the completion of the social media investigation, we then proceeded with instant messaging step. Instant messaging accounts were tied to different mobile numbers. In this experiment, we registered instant messaging account using mobile number "+60162444415" and we performed the communication with mobile number "+60125999159". Table 5 shows all the steps and activities taken for instant messaging experiment

Table 5 - Detail Steps for Instant Messaging Experiment

| Instant Messaging | Steps and Action | Image Filename |
|---|---|---|
| **Telegram Application** | Installing Telegram Applications | Telegram-Install |
|  | Register Telegram account<br>- select Malaysia +60162444415<br>- generating telegram code : 99246<br>- received sms from +93450009276 | Telegram-Register |
|  | Telegram Activities<br>- Sent to +60125999159 : Hi<br>- Received from +60125999159 : Hello<br>- Received from +60125999159 : InstaSize_2015_4_36453.jpg<br>- Save picture - asking for access picture ><br>- Received from +60125999159 : 200507000.pdf<br>- Save file : asking for access memory card > | Telegram-chat |
| **Telegram Web** | Login Telegram thru web web.telegram.org<br>- select Malaysia +60162444415<br>- generating telegram code 34303<br>- received sms from +93450009276 | Telegram-Browse |
|  | Telegram Activities<br>- Sent to +60125999159 : Hi from web<br>- Received from +60125999159 : Hi back<br>- Received from +60125999159 : IMG_20150508_200905.jpg<br>- Save picture : failed<br>- Sent to +60125999159 : Thanks for the picture | Telegram-webxtvt |
| **OpenWapp** | Installing OpenWapp Application | OpenWapp-install |
|  | Register WhatsApp<br>- select Malaysia +60162444415<br>- generating WhatsApp code 560-103<br>- received sms from 63365 | OpenWapp-reg |
|  | OpenWapp Activties<br>- Received from +60125999159 : Hello | OpenWapp-chat |

| | - Sent to +60125999159 : Testing openwapp from Firefox phone<br>- Received from +60125999159 : IMG_20150508_171904.jpg<br>- Save picture : agxahpqv0gm3suwkv7glgnxz6ig4k_bh6jq_bivja5q0.jpg | |
|---|---|---|
| **Line Application** | Installing Line Application | Line-Install |
| | Register Line Account<br>- select Malaysia +60162444415<br>- Reg code 1877<br>- Received sms from +601117224258 | Line-Reg |
| | Line Activities<br>- najwadi added syamsaziela as friend<br>- najwadi sent msg : Hi syamsaziela (6.15PM)<br>- syamsaziela replied : send to firefox phone (6.15PM)<br>- syamsaziela sent : IMG_8933.jpg<br>- najwadi received, saved (read 6.18PM) : New folder name LINE - 2015412_185355518.jpg | Line-xtvt |

In instant messaging experiment, we have performed the acquisition processes 11 times for each internal phone image and memory image. The files were also named according to the step taken as shown in Table 5. The experiment were performed using OpenWapp and Line applications, and both Telegram application and mobile web. OpenWapp is the third party application for WhatsApp in FxOS because WhatsApp was not officially offered in Mozilla Market Place. Both WhatsApp and Line does not offer mobile web support. During our investigation, we restored the phone to their factory setting 4 times, the number of experiment conducted.

## 3.5 Comparing Base Image with Activities Images

Acquired forensic images were analyzed using AccessData Forensic Toolkit (FTK) version 3.1.2 and HxD Hex Editor 1.7.7.0. Our major analysis technique were solely rely on the actions and execution to the phone during experiment. The base image were then compared with the captured forensic images according to the detail steps and actions taken. MD5 hash value of all images were recorded as shows in Table 6.

Table 6 - MD5 Hash Value for Acquired Images

| Filename | Phone Image (.ffp) | Memory Image (.ffm) |
|---|---|---|
| **Base** | F080B51EDBCCA1DFB85023A96C86B95D | 7D93024506D837EB85682AC6C2DAE7A9 |
| **Facebook-Install** | 94A4CCEB5333D5A8D7E8498E531DFCF7 | 24E0273A4CB7C852E20230E823675A50 |
| **Facebook-Login** | 755811F684ED0A987E07C2B27D047560 | DA9A48B9F04F73415BE37E26F23B4612 |
| **Facebook-Post** | 51ABC84775AEA4326814707450D21D23 | 84EDE34F2059CE45CE55C195E6B4F116 |
| **Facebook-Browse** | FCB0C5CFEEDC292185B340C809B63391 | A040D89CBFBBD3F8A908F2E9F9ED4C3E |
| **Facebook-xtvt** | 486719D871E0B71D198099E82C13D959 | 3C3F778E0FA59A93B4274265E4515967 |
| **Twitter-Install** | BF63A51A105A6DDA94FC6A55D511FD76 | E3178BA282723C70DBBB65DE9CBA2FFB |
| **Twitter-Login** | 6B7F1035F1B2C4DECA3B793F3AF36CF5 | 987C447CC2B7651D283B53021121E93F |
| **Twitter-Tweet** | 6A18EDCAD1A2AF26CA274067D1F27B34 | 516FB1B243A47AB1256B121321BF98E6 |
| **Twitter-Browse** | 57F6990C23DF171C32EF8BF81FF550A3 | 242247A35D866D8E9F642D171457EC52 |
| **Twitter-xtvt** | 4CE51367D17FD6FDE29EAC744E763A3B | 872975E4C17319CCD44E0AF7E968CF33 |
| **Google-Login** | C5B877501D82C2CA6998219FF3767FFD | E94A7D2141E3CF18A146AAA8BCB6D181 |
| **Google-xtvt** | 5BCBBD245F29662C7127B20B8341952B | CDD22BA569641831292E85C9C8F860E1 |
| **Telegram-Install** | B15D74872C6E42244F859A73AD24C25B | 079F5A0CBEDB4DD920B90A8CA26E3FCB |
| **Telegram-Register** | A46445F6DB4A79C6FFD354CED983DF0A | A43650BD2947C1CEF6AB1117C7F4070F |
| **Telegram-chat** | 2424B5DE62DBE83F6BB8DAD19753A84D | 55F5E5BD6B575113F3600F3A381F0D16 |
| **Telegram-Browse** | B3DE5669770CEAAF61F602499513328D | 3EB424F3D1D575F3E7A86A31948A83AB |
| **Telegram-webxtvt** | F261B40B276A572C93C3D34AD9738444 | A0E12223ADD6BC1C7845B12527B1581C |
| **OpenWapp-install** | 1BD68D8D6BD787AC41AC67E9CD1726AC | B4F80CF7E711D4FBD595673DB3DE84CB |
| **OpenWapp-reg** | EFE5DB509CEFFA005F4A60337AA3CF2F | BD6F684E93D53DE79E616088165A306E |
| **OpenWapp-chat** | 56590158061E662FCAED3BC1A5C773ED | 517D345BE8FBAD1FFFF45ED5CA7345F5 |
| **Line-Install** | C5E85B3BA980E8DA7130DBB9556C6501 | 29908DE05F3B7DECEA3B17ADF1EDD946 |
| **Line-Reg** | 992818E14DD9684715EACB812CBB6FED | EF641273ABF57E76F4516803DEDE8216 |
| **Line-xtvt** | D13BF7662CB491E73DB6ABF7E193F401 | 56C0CBE3B8DD306EBFFEDA68EEBDF5C3 |

MD5 hash value were taken and preserved. Any modification of forensic images will be easily detected by looking at their hash value. In our experiment, none of these forensic images shared the same hash values. This conclude that our forensic images was giving differences in value, modification and evidence in it.

## 4. Discussion and Analysis

In this section, we have divided our analysis into two parts. The first analysis has focused on social media investigation, while the second part has focused on instant messaging investigation. For both analysis, we started with the internal phone images (.ffp) and then proceeded with memory images (.ffm). The purpose of separating the two analysis was to differentiate between both analysis results. By doing this analysis separately, we have discovered that certain information which was not recoverable in the phone images, was able to be retrieved from memory images.

### 4.1 Social Media Investigation

In social media experiment, we were focusing the investigation for 2 social media applications and 3 social media mobile web. We have successfully separating the captured images based on major steps and details as shown in Table 3. There were several forensic worth of evidence that we were trying to recover and trace. First, we have explored the residual artifacts generated by application and URL involved for mobile web. Second, we were trying find any ID name that able to be captured after we login into social media. Third, we have searched for any credential involving username and password after login process. Forth, we have traced back what activities that has been captured in the images. Lastly we have checked for the data remnant and left over after complete uninstallation of the application.

### 4.1.1 Social Media Phone Image

Each of the phone images were opened using AccessData Forensic Toolkit (FTK) version 3.1.2 and HxD Hex Editor 1.7.7.0. The analysis were conducted manually by searching several keyword related to previous experiment in forensically sounds manner. The first binary search performed was the application keyword. This search executed once the selected application successfully installed from Mozilla Market Place. Since our factory setting base image was only comes with preinstalled application and none of it were used in our investigation, we cannot find any keyword of selected social media from the base image. Figure 2 shows Twitter keyword appear when we executed the first search.

```
23CD4320  63 6F 6E 6E 65 63 74 5F 6A 6F 69 6E 65 64 5F 74   connect_joined_t
23CD4330  77 69 74 74 65 72 5F 6F 6E 65 22 3A 22 59 6F 75   witter_one":"You
23CD4340  72 20 63 6F 6E 74 61 63 74 20 3C 73 70 61 6E 20   r contact <span
23CD4350  63 6C 61 73 73 3D 27 68 69 67 68 6C 69 67 68 74   class='highlight
23CD4360  27 3E 7B 7B 6E 61 6D 65 7D 7D 3C 2F 73 70 61 6E   '>{{name}}</span
23CD4370  3E 20 28 40 7B 7B 73 63 72 65 65 6E 5F 6E 61 6D   > (@{{screen_nam
23CD4380  65 7D 7D 29 20 69 73 20 6F 6E 20 54 77 69 74 74   e}}) is on Twitt
23CD4390  65 72 22 2C 22 63 6F 6E 6E 65 63 74 5F 72 65 74   er","connect_ret
23CD43A0  77 65 65 74 65 64 5F 62 79 5F 6D 61 6E 79 22 3A   weeted_by_many":
23CD43B0  22 3C 73 70 61 6E 20 63 6C 61 73 73 3D 27 68 69   "<span class='hi
23CD43C0  67 68 6C 69 67 68 74 27 3E 7B 7B 6E 61 6D 65 7D   ghlight'>{{name}
23CD43D0  7D 3C 2F 73 70 61 6E 3E 20 61 6E 64 20 3C 73 70   }</span> and <sp
23CD43E0  61 6E 20 63 6C 61 73 73 3D 27 68 69 67 68 6C 69   an class='highli
23CD43F0  67 68 74 27 3E 7B 7B 6E 75 6D 62 65 72 7D 7D 20   ght'>{{number}}
```

Figure 2 - Application Keyword Found After Installation in Phone Image

The second search were performed to find URL for social media mobile web. Mobile web is totally different with application. We do not had to install it but only need to browse using preinstalled Firefox browser. Figure 3 shows Twitter URL was found in the images and the result has also showing the visited date and time.

```
9D3573F0  00 00 00 00 00 00 00 00 00 00 00 00 00 00 00 00   ................
9D357400  00 01 00 13 00 00 00 00 00 00 00 01 55 4B DE 1C   ............UKÞ.
9D357410  55 4B DE 1D FF FF FF FF 00 00 00 00 00 00 00 25   UKÞ.ÿÿÿÿ.......%
9D357420  00 00 15 9D 48 54 54 50 7E 31 30 31 38 7E 31 3A   ....HTTP~1018~1:
9D357430  68 74 74 70 73 3A 2F 2F 77 77 77 2E 74 77 69 74   https://www.twit
9D357440  74 65 72 2E 63 6F 6D 2F 00 72 65 71 75 65 73 74   ter.com/.request
9D357450  2D 6D 65 74 68 6F 64 00 47 45 54 00 72 65 73 70   -method.GET.resp
9D357460  6F 6E 73 65 2D 68 65 61 64 00 48 54 54 50 2F 31   onse-head.HTTP/1
9D357470  2E 31 20 33 30 31 20 4D 6F 76 65 64 20 50 65 72   .1 301 Moved Per
9D357480  6D 61 6E 65 6E 74 6C 79 0D 0A 43 6F 6E 74 65 6E   manently..Conten
9D357490  74 2D 4C 65 6E 67 74 68 3A 20 30 0D 0A 44 61 74   t-Length: 0..Dat
9D3574A0  65 3A 20 54 68 75 2C 20 30 37 20 4D 61 79 20 32   e: Thu, 07 May 2
9D3574B0  30 31 35 20 32 31 3A 31 35 3A 35 31 20 47 4D 54   015 21:15:51 GMT
9D3574C0  0D 0A 4C 6F 63 61 74 69 6F 6E 3A 20 68 74 74 70   ..Location: http
9D3574D0  73 3A 2F 2F 74 77 69 74 74 65 72 2E 63 6F 6D 2F   s://twitter.com/
```

Figure 3 - URL for Social Media Mobile Web Found in Phone Image

The third search were to find our profile name or user ID once we logged into the social media. Profile name is different with username. Username used to login into the social media while profile name were the name displayed in our social media account. Figure 4 shows our profile name together with the user ID and Facebook profile path that we managed to find.

```
61FBE340  30 61 62 34 32 39 62 30 66 32 62 37 66 34 39 39  0ab429b0f2b7f499
61FBE350  38 66 26 6F 65 3D 35 35 43 33 44 32 44 34 26 5F  8f&oe=55C3D2D4&_
61FBE360  5F 67 64 61 5F 5F 3D 31 34 33 38 39 36 34 32 35  _gda__=143896425
61FBE370  38 5F 37 38 38 66 66 32 65 66 36 35 39 63 36 65  8_788ff2ef659c6e
61FBE380  30 66 66 38 38 33 39 37 38 32 32 38 65 66 63 35  0ff883978228efc5
61FBE390  30 65 22 2C 22 74 65 78 74 22 3A 22 52 6F 6F 74  0e","text":"Root
61FBE3A0  20 57 61 64 69 22 2C 22 75 69 64 22 3A 31 30 30   Wadi","uid":100
61FBE3B0  30 30 35 36 37 36 36 30 38 37 30 35 2C 22 70 61  005676608705,"pa
61FBE3C0  74 68 73 22 3A 5B 5D 2C 22 62 6F 6F 74 73 74 72  ths":[],"bootstr
61FBE3D0  61 70 22 3A 31 7D 2C 7B 22 70 61 74 68 22 3A 22  ap":1},{"path":"
61FBE3E0  2F 70 72 6F 66 69 6C 65 2E 70 68 70 3F 69 64 3D  /profile.php?id=
61FBE3F0  37 31 32 31 36 39 30 34 33 22 2C 22 70 68 6F 74  712169043","phot
61FBE400  6F 22 3A 22 68 74 74 70 73 3A 2F 2F 66 62 63 64  o":"https://fbcd
61FBE410  6E 2D 70 72 6F 66 69 6C 65 2D 61 2E 61 6B 61 6D  n-profile-a.akam
```

Figure 4 - Facebook Profile Name Appear in Phone Image.

One of the most valuable information that we need to protect in social media, were the account credentials. In the event that the account credentials fall into wrong hands, they can definitely log in into our account and pretended to be us. When they commit any crime using our account, prosecution will be charged under our name. Next search were to find username and password that we used during our login process. Username rules were different between social media. Facebook and Google+ only accept email as username, while twitter can accept either email or selected string. Figure 5 shows Twitter username appear in our search.

```
23DCDC60  22 66 6F 6C 6C 6F 77 65 64 5F 62 79 22 3A 66 61  "followed_by":fa
23DCDC70  6C 73 65 7D 2C 22 74 6F 6B 65 6E 73 22 3A 5B 7B  lse},"tokens":[{
23DCDC80  22 74 6F 6B 65 6E 22 3A 22 6D 6F 68 64 22 7D 2C  "token":"mohd"},
23DCDC90  7B 22 74 6F 6B 65 6E 22 3A 22 6E 61 6A 77 61 64  {"token":"najwad
23DCDCA0  69 22 7D 2C 7B 22 74 6F 6B 65 6E 22 3A 22 77 61  i"},{"token":"wa
23DCDCB0  64 69 65 71 22 7D 2C 7B 22 74 6F 6B 65 6E 22 3A  dieq"},{"token":
23DCDCC0  22 40 77 61 64 69 65 71 22 7D 5D 2C 22 69 6E 6C  "@wadieq"}],"inl
23DCDCD0  69 6E 65 22 3A 66 61 6C 73 65 2C 22 66 6F 6C 6C  ine":false,"foll
23DCDCE0  6F 77 65 64 5F 62 79 22 3A 66 61 6C 73 65 7D 5D  owed_by":false}]
23DCDCF0  7D 7D 2C 22 74 6F 70 69 63 73 54 79 70 65 61 68  }},"topicsTypeah
23DCDD00  65 61 64 22 3A 7B 22 75 70 64 61 74 65 64 41 74  ead":{"updatedAt
23DCDD10  22 3A 31 34 33 31 30 33 33 36 32 38 32 33 39 2C  ":1431033628239,
23DCDD20  22 64 61 74 61 22 3A 7B 22 61 70 69 22 3A 22 74  "data":{"api":"t
23DCDD30  6F 70 69 63 73 54 79 70 65 61 68 65 61 64 22 2C  opicsTypeahead",
```

Figure 5 - Username in Twitter Phone Image

Activities and communication in social media have contributed a massive valuable information in forensic investigation. From the communication pattern, we might be able to identify if there is any cyber bullying as well as cyber stalking. We also can also investigate if any sexual harassment had occurred. In the next search, we were trying to find activities that can be recovered in our experiment. We purposely used a certain communication string in Malaysian language so that the search will not redundant with common word like "hello" and "hi" in the images. Figure 6 below shows the communication string that we use earlier in Twitter experiment, appearing in the search result.

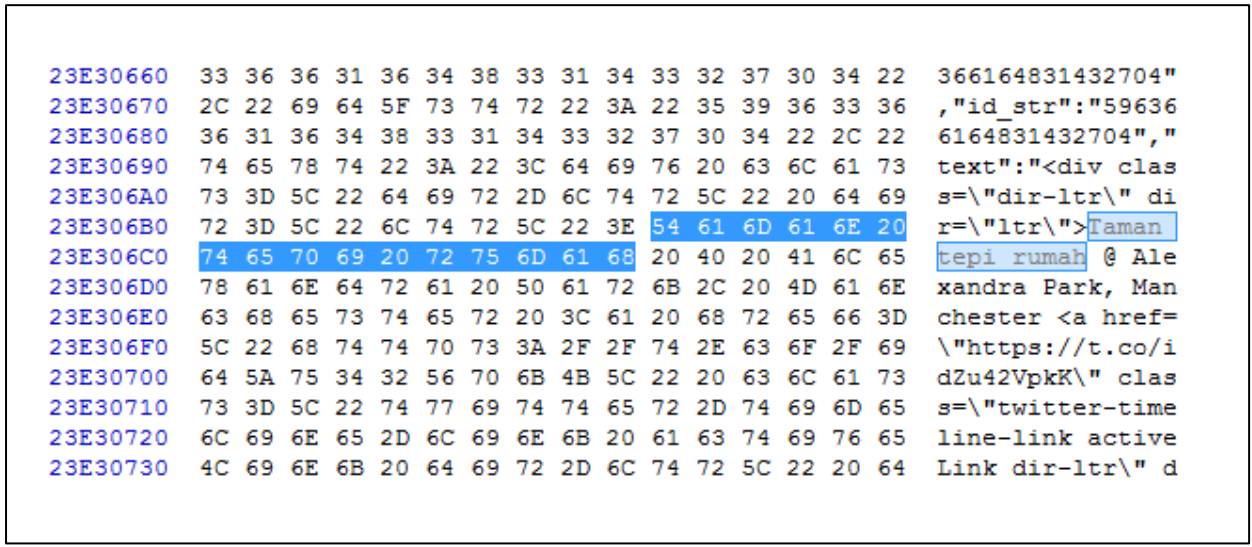

Figure 6 - Communication String That Able to Recover from Phone Image

The search process were executed by using prepared detail steps with the phone images accordingly. If the information were able to be found and recovered, we marked it as right. After all search has been executed and completed, the findings were recorded as shows in Table 7

Table 7 - Social Media Phone Image Finding Summary

| Evidences | Facebook Application | Facebook Web | Twitter Application | Twitter Web | Google+ Web |
|---|---|---|---|---|---|
| Application keyword | ✓ | N/A | ✓ | N/A | N/A |
| Web URL | N/A | ✓ | N/A | ✓ | ✓ |
| Name after login | ✓ | ✓ | ✓ | ✓ | ✗ |
| Username after login | ✗ | ✗ | ✓ | ✓ | ✗ |
| Password after login | ✗ | ✗ | ✗ | ✗ | ✗ |
| Activities | ✗ | ✗ | ✓ | ✓ | ✗ |
| Uninstalled data remnant | ✓ | N/A | ✓ | N/A | N/A |

In general, all application keyword will appear in the search once we successfully installed the application. These result show that, there were changes in application list once it was installed. The same search result happened to web URL keyword. Once we visited the web, the keyword will be stored in the browsing history, including date and time stamp. When we login into the selected social media, all profile names were recorded in the phone except Google+ mobile web. As for social media username, only Twitter application and Twitter mobile web were able for retrieval. No password can be seen for all social media. By the same token, the communications and activities only appear in Twitter application and Twitter mobile web. No activities can be seen in Facebook and Google+. Data remnant were able to be retrieved from both Facebook and Twitter application.

### 4.1.2 Social Media Phone and Memory Images

Analysis of the memory images were the same with the previous analysis. We have also performed the keyword search by using a prepared detail steps. The search were repeated, but confined to only for the value that does not appear in our first search using phone images. Each of the memory images were also opened using AccessData FTK and HxD Hex Editor. For this analysis, we focused more on social media credential and communication activities, since most of this part were not recovered from the phone images analysis. Figure 7 shows Facebook credential that we used to login appearing in our first search attempt.

```
02178E30  00 00 00 00 82 FF FF FF 00 00 00 00 82 FF FF FF  ....,ÿÿÿ....,ÿÿÿ
02178E40  00 00 00 00 82 FF FF FF 00 00 00 00 87 FF FF FF  ....,ÿÿÿ....‡ÿÿÿ
02178E50  58 03 1C 0C 80 36 16 0E 00 00 00 00 70 DE 1E 0E  X...€6......pÞ..
02178E60  06 00 00 00 37 00 00 00 B0 5F A5 0F 00 00 00 00  ....7...°_¥.....
02178E70  6D 6F 68 64 2E 6E 61 6A 77 61 64 69 40 67 6D 61  mohd.najwadi@gma
02178E80  69 6C 2E 63 6F 6D 00 6D 6F 68 64 2E 6E 61 6A 77  il.com.mohd.najw
02178E90  61 64 69 40 67 6D 61 69 6C 2E 63 6F 6D 00 6E 61  adi@gmail.com.na
02178EA0  6A 77 61 64 69 38 37 FF 00 00 00 00 82 FF FF FF  jwadi87ÿ....,ÿÿÿ
02178EB0  00 00 00 00 82 FF FF FF 00 00 00 00 82 FF FF FF  ....,ÿÿÿ....,ÿÿÿ
02178EC0  00 00 00 00 82 FF FF FF 00 00 00 00 82 FF FF FF  ....,ÿÿÿ....,ÿÿÿ
02178ED0  00 00 00 00 82 FF FF FF 00 00 00 00 87 FF FF FF  ....,ÿÿÿ....‡ÿÿÿ
02178EE0  58 03 1C 0C 80 36 16 0E 00 00 00 00 DF 1E 0E     X...€6.......ß..
02178EF0  0B 00 00 00 59 00 00 00 10 70 A5 0F 00 00 00 00  ....Y....p¥.....
02178F00  41 55 54 48 20 50 4C 41 49 4E 20 62 57 39 6F 5A  AUTH PLAIN bW9oZ
```

Figure 7 - Facebook Credential Found Without Encrypted in Memory Image

The search was continued with other social media credential and communication activities. The summary in Table 8 show the result of what evidences can be retrieved from both phone images and memory images.

Table 8 - Social Media Phone Images and Memory Images Finding Summary

| Evidences | Facebook Application | Facebook Web | Twitter Application | Twitter Web | Google+ Web |
|---|---|---|---|---|---|
| **Keyword after install** | ✓ | N/A | ✓ | N/A | N/A |
| **Web URL** | N/A | ✓ | N/A | ✓ | ✓ |
| **Name after login** | ✓ | ✓ | ✓ | ✓ | ✓ |
| **Username after login** | ✓ | ✓ | ✓ | ✓ | ✗ |
| **Password after login** | ✓ | ✓ | ✓ | ✓ | ✗ |
| **Activities** | ✗ | ✗ | ✓ | ✓ | ✗ |
| **Uninstalled data remnant** | ✓ | N/A | ✓ | N/A | N/A |

The finding of this analysis has shown that, all credentials especially password cannot be seen in phone images, but we were able to retrieve it in memory images except for the Google+. As for the communication activities, we still do not find any keyword matching in Facebook and Google+. We then randomly checked with Facebook and Google+ memory images to find any communication occurrence other than what we recorded, but failed to find any. The result of the study was suggesting that the communication activities in Facebook and Google+ for FxOS were either encrypted or stored in their respective server only.

## 4.2 Instant Messaging Investigation

For instant messaging experiment and investigations, we have managed to acquire 3 instant messaging applications and 1 instant messaging mobile web activities. First, we have explored the residual artifacts generated by application and URL involved for mobile web. Second, we have tried to find the user phone number used during registration. Third, we have tried to get the registration code and SMS or call verification received. Forth, we have traced back what activities that has been captured in the images. Lastly, we have checked for data remnant after we complete the uninstallation the application.

### 4.2.1 Instant Messaging Phone Image

Like the previous experiment, each of the phone images were opened using AccessData Forensic Toolkit (FTK) version 3.1.2 and HxD Hex Editor 1.7.7.0. The analysis has also conducted by manually searching several keyword related to previous experiment in forensically sounds manner. The first binary search performed was the application keyword. This search executed once the selected application successfully installed from Mozilla Market Place. Since our factory setting base image was only comes with preinstalled application and none of it were used in our investigation, we cannot find any keyword of selected instant messaging from the base image. Figure 8 shows Telegram keyword appear when we executed the first search.

```
24101620  7C A2 35 C7 81 2D C0 FC 9F C6 21 41 1C 5F A9 0D  |¢5Ç.-ÀüŸ!A._©.
24101630  EB 4F BE B6 01 F2 7F 3C 3F 4F 32 14 01 B8 76 0E  ëO¾¶.ò.<?O2..¸v.
24101640  4B 94 FD 03 87 CB 18 07 52 FF 71 3E 0F B7 28 F8  K"ý.‡Ë..Rÿq>.·(ø
24101650  84 56 73 20 22 7F C7 81 1F D8 4B C3 9C 7F 10 27  „Vs ".Ç..ØKÃœ..'
24101660  D8 D5 1B FC 0F AE 91 FF 00 50 4B 03 04 14 00 00  ØÕ.ü.®'ÿ.PK....
24101670  00 08 00 29 8C 4D 46 ED 78 7C 96 43 04 00 00 7B  ...)ŒMFíx|-C...{
24101680  05 00 00 10 00 18 00 69 6D 67 2F 54 65 6C 65 67  .......img/Teleg
24101690  72 61 6D 2E 70 6E 67 55 54 05 00 03 2E 0B DE 54  ram.pngUT.....ÞT
241016A0  75 78 0B 00 01 04 F6 01 00 00 04 14 00 00 00 EB  ux....ö........ë
241016B0  0C F0 73 E7 E5 92 E2 62 60 60 E0 F5 F4 70 09 02  .ðsçå'âb``àõôp..
241016C0  D2 F6 40 2C C0 C1 06 24 17 2E 5D 6E 03 A4 24 4B  Òö@,ÀÁ.$..]n.¤$K
241016D0  5C 23 4A 82 F3 D3 4A CA 13 8B 52 19 1C 53 F2 93  \#J‚óÓJÊ.‹R..Sò"
241016E0  52 15 3C 73 13 D3 53 83 52 13 53 2A 0B 4F A6 02  R.<s.ÓSƒR.S*.O¦.
241016F0  15 31 2B 65 86 44 94 44 F8 FA 58 25 E7 E7 EA 25  .1+e†D"DøúX%ççê%
```

Figure 8 - Application keyword Found After Installation in Phone Memory

The second search were executed to find URL for instant messaging mobile web, in this experiment only Telegram we tested for mobile web. Mobile web is totally different with the application. We do not had to install, instead we only need to browse using preinstalled Firefox browser. Figure 9 shows Telegram URL was found in the images and the result was also showing the visited date and time.

```
9CB253D0  00 00 00 00 00 00 00 00 00 00 00 00 00 00 00 00  ................
9CB253E0  00 00 00 00 00 00 00 00 00 00 00 00 00 00 00 00  ................
9CB253F0  00 00 00 00 00 00 00 00 00 00 00 00 00 00 00 00  ................
9CB25400  00 01 00 13 B1 00 00 35 00 00 00 03 55 4D 16 68  ....±..5....UM.h
9CB25410  55 4D 16 69 55 4D 1C 5C 00 00 02 7D 00 00 00 26  UM.iUM.\...}...&
9CB25420  00 00 10 90 48 54 54 50 7E 31 30 31 38 7E 31 3A  ....HTTP~1018~1:
9CB25430  68 74 74 70 73 3A 2F 2F 77 65 62 2E 74 65 6C 65  https://web.tele
9CB25440  67 72 61 6D 2E 6F 72 67 2F 00 72 65 71 75 65 73  gram.org/.reques
9CB25450  74 2D 6D 65 74 68 6F 64 00 47 45 54 00 72 65 73  t-method.GET.res
9CB25460  70 6F 6E 73 65 2D 68 65 61 64 00 48 54 54 50 2F  ponse-head.HTTP/
9CB25470  31 2E 31 20 32 30 30 20 4F 4B 0D 0A 53 65 72 76  1.1 200 OK..Serv
9CB25480  65 72 3A 20 6E 67 69 6E 78 2F 31 2E 36 2E 32 0D  er: nginx/1.6.2.
9CB25490  0A 44 61 74 65 3A 20 46 72 69 2C 20 30 38 20 4D  .Date: Fri, 08 M
9CB254A0  61 79 20 32 30 31 35 20 31 39 3A 32 38 3A 31 32  ay 2015 19:28:12
9CB254B0  20 47 4D 54 0D 0A 43 6F 6E 74 65 6E 74 2D 54 79   GMT..Content-Ty
```

Figure 9 - URL for Instant Messaging Mobile Web Found in Phone Image

The third search were to find the phone number used during registration. Most of our selected instant messaging services asked the phone number to be tied with account registration. Instant messaging services like Telegram and WhatsApp will display the phone number together with selected name when we send private message to other user, while Line will not display any number but only selected name. The confirmation during registration were either code from SMS or received call from the providers. Figure 10 and Figure 11 shows the phone number and registration code for WhatsApp services respectively.

```
A59787E0  00 00 00 00 00 00 00 00 00 00 00 00 00 00 00 00  ................
A59787F0  00 00 00 00 00 00 00 00 00 00 00 00 00 00 00 00  ................
A5978800  7B 22 73 74 61 74 75 73 22 3A 22 6F 6B 22 2C 22  {"status":"ok","
A5978810  6C 6F 67 69 6E 22 3A 22 36 30 31 36 32 34 34 34  login":"60162444
A5978820  34 31 35 22 2C 22 70 77 22 3A 22 4E 43 69 30 66  415","pw":"NCi0f
A5978830  45 50 6F 51 43 69 37 59 55 50 65 63 53 64 30 63  EPoQCi7YUPecSd0c
A5978840  61 38 71 5A 51 41 3D 22 2C 22 74 79 70 65 22 3A  a8qZQA=","type":
A5978850  22 65 78 69 73 74 69 6E 67 22 2C 22 65 78 70 69  "existing","expi
A5978860  72 61 74 69 6F 6E 22 3A 31 34 36 32 36 30 37 39  ration":14626079
A5978870  39 36 2C 22 6B 69 6E 64 22 3A 22 66 72 65 65 22  96,"kind":"free"
A5978880  2C 22 70 72 69 63 65 22 3A 22 55 53 24 30 2E 39  ,"price":"US$0.9
A5978890  39 22 2C 22 63 6F 73 74 22 3A 22 30 2E 39 39 22  9","cost":"0.99"
A59788A0  2C 22 63 75 72 72 65 6E 63 79 22 3A 22 55 53 44  ,"currency":"USD
A59788B0  22 2C 22 70 72 69 63 65 5F 65 78 70 69 72 61 74  ","price_expirat
A59788C0  69 6F 6E 22 3A 31 34 33 33 37 39 38 31 39 31 7D  ion":1433798191}
```

Figure 10 - Phone Number Used During Registration Found in Phone Memory

```
9CBC1420  00 00 10 3C 48 54 54 50 7E 31 30 32 31 7E 30 3A  ...<HTTP~1021~0:
9CBC1430  68 74 74 70 73 3A 2F 2F 76 2E 77 68 61 74 73 61  https://v.whatsa
9CBC1440  70 70 2E 6E 65 74 2F 76 32 2F 72 65 67 69 73 74  pp.net/v2/regist
9CBC1450  65 72 3F 63 63 3D 36 30 26 69 6E 3D 31 36 32 34  er?cc=60&in=1624
9CBC1460  34 34 34 31 35 26 63 6F 64 65 3D 35 36 30 31 30  44415&code=56010
9CBC1470  33 26 69 64 3D 25 42 44 25 41 45 25 32 31 25 46  3&id=%BD%AE%21%F
9CBC1480  39 25 44 31 25 45 41 25 36 44 25 36 32 25 35 38  9%D1%EA%6D%62%58
9CBC1490  25 46 35 25 45 41 25 34 42 25 38 31 25 36 33 25  %F5%EA%4B%81%63%
9CBC14A0  45 33 25 36 39 25 39 46 25 44 43 25 45 37 25 34  E3%69%9F%DC%E7%4
9CBC14B0  38 00 72 65 71 75 65 73 74 2D 6D 65 74 68 6F 64  8.request-method
9CBC14C0  00 47 45 54 00 72 65 73 70 6F 6E 73 65 2D 68 65  .GET.response-he
9CBC14D0  61 64 00 48 54 54 50 2F 31 2E 31 20 32 30 30 20  ad.HTTP/1.1 200
9CBC14E0  4F 4B 0D 0A 53 65 72 76 65 72 3A 20 59 61 77 73  OK..Server: Yaws
9CBC14F0  20 31 2E 39 38 0D 0A 44 61 74 65 3A 20 46 72 69  1.98..Date: Fri
9CBC1500  2C 20 30 38 20 4D 61 79 20 32 30 31 35 20 32 32  , 08 May 2015 22
```

Figure 11 - Registration Code Received During Account Confirmation Found in Phone Memory

The search process were executed by using a prepared detail steps with the phone images accordingly. If the information were able to be found and recovered, we marked it as right. After all search has been executed and completed, the findings were recorded as shows in Table 9

Table 9 - Instant Messaging Phone Image Finding Summary

| Evidences | Telegram Application | Telegram Web | OpenWapp Application | Line Application |
|---|---|---|---|---|

| | | | | | |
|---|---|---|---|---|---|
| **Keyword after install** | ✓ | N/A | ✓ | ✓ | |
| **Web URL** | N/A | ✓ | N/A | N/A | |
| **Number after register** | ✗ | ✗ | ✓ | ✗ | |
| **Registration code** | ✓ | ✓ | ✓ | ✓ | |
| **SMS/Call verification** | ✓ | ✓ | ✓ | ✓ | |
| **Activities** | ✗ | ✗ | ✗ | ✗ | |
| **Uninstalled data remnant** | ✓ | N/A | ✓ | ✓ | |

In general, all application keyword will appear in the search once we have successfully installed the application. These result shows that, there were changes in application list once it was installed. The same search result happened to web URL keyword. Once we visited the web, the keyword were stored in the browsing history, including date and time stamp. We try to find the phone number we were used during registration but we only manage to find it in OpenWapp phone images. However, all registration code, SMS as well as call verification were manage to find in all instant messaging phone images. In contrary, no communication activities were manage to trace in this analysis. Data remnant were able to be retrieved from both Facebook and Twitter application.

### 4.2.2 Instant Messaging Phone and Memory Images

Analysis of the memory images were the same with the previous analysis. We have also performed the keyword search using a prepared detail steps. The search were repeated, but confined to only for the value that does not appear in our first search using phone images. Each of the memory images were also opened using AccessData Forensic Toolkit (FTK) version 3.1.2 and HxD Hex Editor 1.7.7.0. For this analysis, we focused more on registration number and instant messaging communication activities, since most of this part were not recovered from the phone images analysis. Figure 12 shows OpenWapp communication string that we used during this experiment.

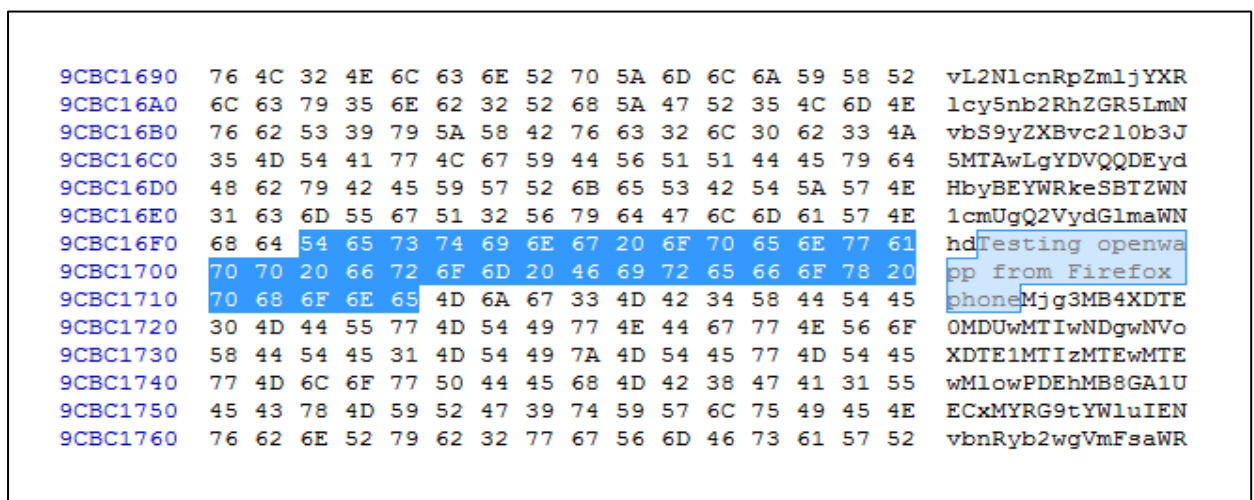

Figure 12 - OpenWapp Communication String Found in Memory Image

The search were continued with other instant messaging communication activities and the phone number used in Telegram and Line. The summary in Table 9 shows the result of what evidences can be retrieved from both phone images and memory images.

Table 9 - Instant Messaging Phone Images and Memory Images Finding Summary

| Evidences | Telegram Application | Telegram Web | OpenWapp Application | Line Application |
|---|---|---|---|---|
| **Keyword after install** | ✓ | N/A | ✓ | ✓ |
| **Web URL** | N/A | ✓ | N/A | N/A |
| **Number after register** | ✓ | ✓ | ✓ | ✓ |
| **Registration code** | ✓ | ✓ | ✓ | ✓ |
| **SMS/Call verification** | ✓ | ✓ | ✓ | ✓ |
| **Activities** | ✗ | ✗ | ✓ | ✗ |
| **Uninstalled data remnant** | ✓ | N/A | ✓ | ✓ |

The finding of this analysis has shown that, all registration phone numbers that were not able to be seen in phone images, were able to be traced in memory images. As for the communication activities, we have only manage to find the communication string in OpenWapp and still do not find any keyword matching in Telegram and Line. We then randomly checked with Telegram and Line memory images to find any communication occurrence other than what we recorded, but also failed to find any. The result of the studies suggested that the communication activities in Telegram and Line for FxOS were either encrypted or stored in their respective server only.

## 6. Conclusion

In this research, we have successfully acquired 24 forensic images, each from FxOS internal phone and volatile memory. The acquired images were then extracted based on the action performed as per documented in detail steps, and were named accordingly. These images were then analyzed and the result was presented and tabled. The result of this research has indicated that most of valuable forensic information were residing in volatile memory. The finding of this study also suggested that memory in FxOS phone were not encrypted, hence it was readable with our forensic tools. Therefore, we managed to recover and trace social media account credential especially on Facebook and Twitter services. In contrary, all untraced information in phone images such as profile name on Google+ and phone number used during registration for Telegram and Line, were traceable in memory images.

The other conclusion drawn from this research were, FxOS applications behaving the same way of its counterpart in mobile web. Therefore, we managed to get the exact same forensic traces and evidences when we analyzed the same services, both in application and mobile web platform. For example Facebook, Twitter and Telegram applications are producing the same forensic trace with its mobile web. In short, these findings have significantly enhanced our understanding of the similarity of FxOS application which design based on WebAPI, with its mobile web platforms. By using WebAPI, FxOS application can executed under very minimal memory requirement, just like opening the web browser.

On the other hand, the analysis of the acquired images has shown that application artifacts were remained in FxOS device after it has been uninstalled. The data remnant and left over from the application folders and browser history will be value added to the investigation and many more information have been successfully retrieved. Further, the findings from this research have also make several contributions to the current set of mobile forensic investigation standard. It is recommend that further research to be undertaken with more broad application and mobile web platform. More forensic investigation of FxOS application would help us to establish greater degree of accuracy in this study.

## References

Accessdata, 2007. Forensic Toolkit (FTK).

Al Mutawa, N., Baggili, I., Marrington, A., 2012. Forensic analysis of social networking applications on mobile devices. Digit. Investig. 9, S24–S33. doi:10.1016/j.diin.2012.05.007

Anglano, C., 2014. Forensic analysis of WhatsApp Messenger on Android smartphones. Digit. Investig. 11, 1–13. doi:10.1016/j.diin.2014.04.003

Chen, S., Hao, X., Luo, M., 2009. Research of Mobile Forensic Software System Based on Windows Mobile, in: 2009 International Conference on Wireless Networks and Information Systems. IEEE, pp. 366–369. doi:10.1109/WNIS.2009.32

Damshenas, M., Dehghantanha, A., Mahmoud, R., 2014. A Survey on Digital Forensics Trends. Int. J. Cyber-Security Digit. Forensics 3, 1–26.

Dezfouli, F.N., Dehghantanha, A., Eterovic-Soric, B., Choo, K.-K.R., 2015. Investigating Social Networking applications on smartphones detecting Facebook, Twitter, LinkedIn and Google+ artefacts on Android and iOS platforms. Aust. J. Forensic Sci. 0618, 1–20. doi:10.1080/00450618.2015.1066854

Dezfouli, F.N., Dehghantanha, A., Mahmod, R., Mohd Sani, N.F., Shamsuddin, S., 2013. A Data-centric Model for Smartphone Security. Int. J. Adv. Comput. Technol. 5, 9–17. doi:10.4156/ijact.vol5.issue9.2

Geeksphone, 2013. Geeksphone Peak [WWW Document]. URL http://www.geeksphone.com/other-devices-2/

Gómez-Miralles, L., Arnedo-Moreno, J., 2012. Versatile iPad forensic acquisition using the Apple Camera Connection Kit. Comput. Math. with Appl. 63, 544–553. doi:10.1016/j.camwa.2011.09.053

Husain, M.I., Sridhar, R., 2010. iForensics : Forensic Analysis of Instant Messaging on. Lect. Notes Inst. Comput. Sci. Soc. Informatics Telecommun. Eng. - Digit. Forensics Cyber Crime 31, 9–18. doi:10.1007/978-3-642-11534-9_2

Iqbal, A., Marrington, A., Baggili, I., 2014. Forensic artifacts of the ChatON Instant Messaging application. Int. Work. Syst. Approaches Digit. Forensics Eng., SADFE. doi:10.1109/SADFE.2013.6911538

Iqbal, B., Iqbal, A., Obaidli, H. Al, 2012. A novel method of iDevice (iPhone, iPad, iPod) forensics without jailbreaking, in: 2012 International Conference on Innovations in

Information Technology (IIT). IEEE, pp. 238–243. doi:10.1109/INNOVATIONS.2012.6207740

Irwin, D., Hunt, R., 2009. Forensic information acquisition in mobile networks, in: 2009 IEEE Pacific Rim Conference on Communications, Computers and Signal Processing. IEEE, pp. 163–168. doi:10.1109/PACRIM.2009.5291378

Jung, J., Jeong, C., Byun, K., Lee, S., 2011. Sensitive Privacy Data Acquisition in the iPhone for Digital Forensic Analysis. Commun. Comput. Inf. Sci. - Secur. Trust Comput. Data Manag. Appl. 186, 172–186. doi:10.1007/978-3-642-22339-6_21

Karpisek, F., Baggili, I., Breitinger, F., 2015. WhatsApp network forensics: Decrypting and understanding the WhatsApp call signaling messages. Digit. Investig. 1–9. doi:10.1016/j.diin.2015.09.002

Lessard, J., Kessler, G.C., 2010. Android Forensics : Simplifying Cell Phone Examinations. Small Scale Digit. Device Forensics J. 4, 1–12.

Maël Hörz, 2009. HxD Hex Editor.

Mohtasebi, S., Dehghantanha, A., 2013. Towards a Unified Forensic Investigation Framework of Smartphones. Int. J. Comput. Theory Eng. 5, 351–355. doi:10.7763/IJCTE.2013.V5.708

Mohtasebi, S., Dehghantanha, A., 2011a. Defusing the Hazards of Social Network Services. Int. J. Digit. Inf. Wirel. Commun. 1, 504–516.

Mohtasebi, S., Dehghantanha, A., 2011b. A Mitigation Approach to the Privacy and Malware Threats of Social Network Services, in: Digital Information Processing and Communications. pp. 448–459. doi:10.1007/978-3-642-22410-2_39

Mohtasebi, S., Dehghantanha, A., Broujerdi, H.G., 2012. Smartphone Forensics : A Case Study with Nokia E5-00 Mobile Phone. Int. J. Digit. Inf. Wirel. Commun. 1, 651–655.

Pooters, I., 2010. Full user data acquisition from Symbian smart phones. Digit. Investig. 6, 125–135. doi:10.1016/j.diin.2010.01.001

QtADB, 2011. QtADB Android Manager [WWW Document]. URL http://qtadb.wordpress.com

Said, H., Yousif, A., Humaid, H., 2011. IPhone forensics techniques and crime investigation, in: The 2011 International Conference and Workshop on Current Trends in Information Technology (CTIT 11). IEEE, pp. 120–125. doi:10.1109/CTIT.2011.6107946

Sylve, J., Case, A., Marziale, L., Richard, G.G., 2012. Acquisition and analysis of volatile memory from android devices. Digit. Investig. 8, 175–184. doi:10.1016/j.diin.2011.10.003

Taylor, M., Hughes, G., Haggerty, J., Gresty, D., Almond, P., 2012. Digital evidence from mobile telephone applications. Comput. Law Secur. Rev. 28, 335–339. doi:10.1016/j.clsr.2012.03.006

Tso, Y.-C., Wang, S.-J., Huang, C.-T., Wang, W.-J., 2012. iPhone social networking for evidence investigations using iTunes forensics, in: Proceedings of the 6th International Conference on Ubiquitous Information Management and Communication - ICUIMC '12. ACM Press, New York, New York, USA, p. 1. doi:10.1145/2184751.2184827

Walnycky, D., Baggili, I., Marrington, A., Moore, J., Breitinger, F., 2015. Network and device forensic analysis of Android social-messaging applications. Digit. Investig. 14, S77–S84. doi:10.1016/j.diin.2015.05.009

Yang, T.Y., Dehghantanha, A., Choo, K.-K.R., Muda, Z., 2016. Windows Instant Messaging App Forensics: Facebook and Skype as Case Studies. PLoS One 11, e0150300. doi:10.1371/journal.pone.0150300

Yusoff, M.N., Mahmod, R., Abdullah, M.T., Dehghantanha, A., 2014a. Mobile Forensic Data Acquisition in Firefox OS, in: The Third International Conference on Cyber Security, Cyber Warfare, and Digital Forensic (CyberSec2014). pp. 27–31.

Yusoff, M.N., Mahmod, R., Abdullah, M.T., Dehghantanha, A., 2014b. Performance Measurement for Mobile Forensic Data Acquisition in Firefox OS. Int. J. Cyber-Security Digit. Forensics 3, 130–140.

Yusoff, M.N., Mahmod, R., Dehghantanha, A., Abdullah, M.T., 2014c. An Approach for Forensic Investigation in Firefox OS, in: The Third International Conference on Cyber Security, Cyber Warfare, and Digital Forensic (CyberSec2014). IEEE, pp. 22–26.

Yusoff, M.N., Mahmod, R., Dehghantanha, A., Abdullah, M.T., 2014d. Advances of Mobile Forensic Procedures in Firefox OS. Int. J. Cyber-Security Digit. Forensics 3, 141–157.